\documentclass[conference,a4paper]{APSIPA2025}
\usepackage{amsmath}
\usepackage{graphicx}
\usepackage{multirow}
\usepackage{threeparttable}
\usepackage[backend=biber,style=ieee,]{biblatex}
\usepackage{cleveref,booktabs}
\usepackage{tabularx}

\crefname{figure}{Fig.}{Figs.}
\Crefname{figure}{Figure}{Figures}
\crefname{equation}{Eq.}{Eqs.}
\Crefname{equation}{Equation}{Equations}
\crefname{table}{Tab.}{Tabs.}
\Crefname{table}{Table}{Tables}
\crefname{section}{Section}{Sections}
\Crefname{section}{Section}{Sections}

\usepackage{geometry}
\usepackage{fancyhdr}

\fancypagestyle{firststyle}{
  \fancyhf{}
  \fancyhead[C]{2025 Asia Pacific Signal and Information Processing Association Annual Summit and Conference (APSIPA ASC)}
}

\usepackage{amssymb}
\def\R{\mathbb{R}}
\usepackage{bm}
\usepackage{subcaption}
\usepackage{siunitx}
\usepackage{url}
\usepackage{amsthm}
\newtheorem{definition}{Definition}

\begin{document}

\title{Drum-to-Vocal Percussion Sound Conversion and \\Its Evaluation Methodology}

\author{
\authorblockN{
Rinka Nobukawa\authorrefmark{1}\authorrefmark{2}\authorrefmark{4},
Makito Kitamura\authorrefmark{1},
Tomohiko Nakamura\authorrefmark{2},
Shinnosuke Takamichi\authorrefmark{3}\authorrefmark{1}\authorrefmark{2} and
Hiroshi Saruwatari\authorrefmark{1}
}

\authorblockA{
\authorrefmark{1}
Graduate School of Information Science and Technology, University of Tokyo, Tokyo, Japan\\
\authorrefmark{2}
National Institute of Advanced Industrial Science and Technology (AIST), Tokyo, Japan\\
\authorrefmark{3}
Faculty of Science and Technology, Keio University, Yokohama, Japan\\
\authorrefmark{4}
E-mail: rinka-nobukawa@g.ecc.u-tokyo.ac.jp\\
}
}

\maketitle
\thispagestyle{firststyle}
\pagestyle{empty}

\begin{abstract}
This paper defines the novel task of drum-to-vocal percussion (VP) sound conversion.
VP imitates percussion instruments through human vocalization and is frequently employed in contemporary a cappella music.
It exhibits acoustic properties distinct from speech and singing (e.g., aperiodicity, noisy transients, and the absence of linguistic structure), making conventional speech or singing synthesis methods unsuitable.
We thus formulate VP synthesis as a timbre transfer problem from drum sounds, leveraging their rhythmic and timbral correspondence.
To support this formulation, we define three requirements for successful conversion: \textit{rhythmic fidelity}, \textit{timbral consistency}, and \textit{naturalness as VP}. We also propose corresponding subjective evaluation criteria.
We implement two baseline conversion methods using a neural audio synthesizer, the real-time audio variational autoencoder (RAVE), with and without vector quantization (VQ).
Subjective experiments show that both methods produce plausible VP outputs, with the VQ-based RAVE model yielding more consistent conversion.

\end{abstract}

\section{Introduction}
Vocal percussion (VP) is a vocal technique that emulates percussive instrument sounds using articulations of the human vocal tract.
It has a cultural background, having been used as a means of transmitting drum sounds in drum dance in West Africa \cite{alma995113713502466} and as a method of teaching the playing of tabla drums in Northern India \cite{patel2003acoustic,atherton2007rhythm}.
VP is also widely employed in contemporary a cappella music, where any style of contemporary music is performed using only the human voice and/or body~\cite{book}.
To computationally handle contemporary a cappella music, the analysis and synthesis of VP sounds is essential due to its central role in rhythm and timbre reproduction.
In the context of music information retrieval, a few studies on VP sound analysis have been conducted; for example, rhythm-centric music information retrieval~\cite{Kapur2004ISMIR}, VP sound classification~\cite{Evain2021BSPC}, and music generation~\cite{Hipke2014AVI}.
However, the synthesis aspect remains largely underexplored, despite its considerable potential to support practice and arrangement, particularly for novice vocalists.
We thus address the problem of VP sound synthesis in this paper.

Tackling the VP sound synthesis problem poses two major challenges.  
The first challenge is how to synthesize VP sounds.  
For voice parts with lyrics, we have previously proposed a method that synthesizes multi-part singing voices by allowing each part to refer the scores of the others~\cite{hyodo24slt_chorus}.  
This approach enhances inter-part synchrony and improves the naturalness of the synthesized ensemble singing voices.
While such a method is effective for language-bearing vocal parts, VP does not convey linguistic contents and functions as a vocal surrogate for percussive instruments.
It often involves articulatory movements outside the standard phonemic inventory, such as unvoiced trills, ingressive airstreams, and fricative bursts~\cite{blaylock17_interspeech}.
These features are not adequately captured by conventional phoneme-based synthesis methods.
Therefore, VP sound synthesis requires a distinct modeling approach beyond standard speech synthesis.
The second challenge is how to evaluate synthesized VP sounds.
Conventional evaluation methods for synthesized audio include the mean opinion score~\cite{cooper2025good} for naturalness, mel cepstral distortion~\cite{kubichek1993mel} for spectral similarity, and metrics such as Fr\'{e}chet audio distance and inception score~\cite{vinay2022evaluating} for generative audio quality.
However, these evaluation methods are not directly applicable to VP synthesis.
Unlike speech or melodic music signals, VP events are typically short, aperiodic, and spectrally noisy, making conventional metrics poorly suited to capture their perceptual qualities.
To accurately assess the quality of synthesized VP sounds, a dedicated evaluation framework that accounts for these perceptual attributes is therefore necessary.

\begin{figure}
    \centering
    \includegraphics[width=0.86\linewidth]{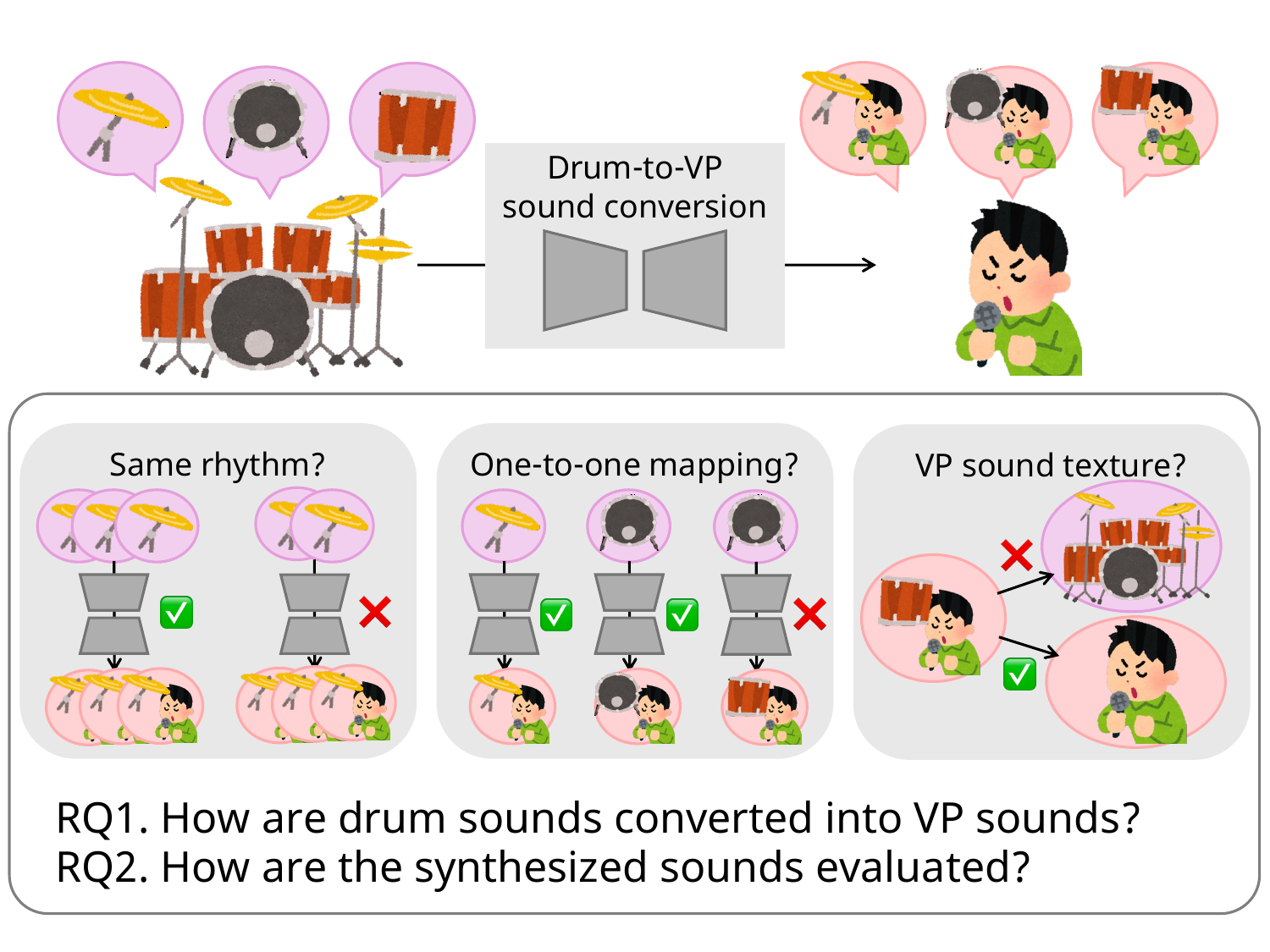}
    \caption{Conceptual illustration of drum-to-VP sound conversion. ``RQ'' denotes a research question.}
    \label{fig:concept}
\end{figure}

To address the aforementioned challenges in synthesis and evaluation, we define a new task: the drum-to-VP conversion task, which aims to convert drum sounds into the corresponding VP renditions (see \cref{fig:concept}).
In this task, we frame the VP synthesis of our interest not as a conventional speech synthesis problem but as a timbre transfer problem from drum sounds to VP sounds. This framing is expected to enable natural acoustic transformations while maintaining rhythmic consistency.
As a baseline model, we adopt the real-time audio variational autoencoder (RAVE)~\cite{RAVE}, a generative model capable of low-latency waveform conversion. In our setting, RAVE is trained on VP sounds and applied to transform input drum signals into their corresponding VP renditions. 
Furthermore, we introduce a dedicated evaluation framework tailored to VP synthesis. This framework consists of three perceptual criteria derived from requirements for our desired timbre transfer: \textit{rhythmic fidelity}, \textit{timbral consistency}, and \textit{naturalness as vocal percussion}. These criteria inform the design of a structured subjective evaluation procedure for the drum-to-VP synthesis. 

\section{Related Works}
\subsection{Vocal Percussion and Human Beatboxing}
VP and human beatboxing are often conflated but have distinct functional and cultural contexts. While beatboxing is typically a solo performance technique that involves a broad range of sound effects, including non-percussive sounds, VP refers to mouth-generated percussive sounds performed in ensemble settings such as contemporary a cappella groups~\cite{hbb}.
In these contexts, VP serves as a substitute for drums and is characterized by its faithful imitation of percussive patterns.  

A guidebook on contemporary a cappella singing~\cite{book} encourages vocal drummers, like their instrumental counterparts, to practice at a tempo where they can execute beats comfortably and accurately.
This highlights the expectation that VP requires rhythmic control similar to that of real drums.  
Given this close functional relationship, drum-to-VP conversion can be regarded as a timbre transfer task rather than a symbolic-to-audio synthesis problem.



\subsection{Acoustic Characteristics of Vocal Percussion}\label{sec:relatedworks}
Several acoustic analyses of VP have been conducted, particularly in relation to pronunciation corresponding to percussion among beatboxing~\cite{blaylock17_interspeech,Proctor2,Paroni2021JASA,Tyte}. VP sounds are typically generated through the vocal tract but differ from conventional speech in their articulatory and acoustic properties. Many VP sounds are aperiodic, unvoiced, and rich in turbulent noise components~\cite{blaylock17_interspeech}. 
This fact suggests prioritizing the magnitude spectrogram over the phase for VP sound synthesis.

Although some VP sounds can be transcribed using the international phonetic alphabet (IPA)~\cite{Proctor2}, especially those produced by less experienced performers, the most skilled VP sounds often defy phonetic categorization~\cite{blaylock17_interspeech,Paroni2021JASA,Tyte}. 
In such cases, articulatory descriptions highlight intentional deviations from speech-like gestures. These findings suggest that phoneme-conditioned synthesis is insufficient and support the use of audio-to-audio models to capture the non-linguistic, instrument-like nature of VP.

\section{Proposed Task: Drum-to-VP Sound Conversion}
\subsection{Task Definition} \label{sec:req}



In this section, we frame the VP sound synthesis task as a problem of converting drum sounds into VP sounds, i.e., the drum-to-VP sound conversion task.  
As discussed in \cref{sec:relatedworks}, VP serves as a vocal surrogate for drums in contemporary a cappella music and often imitates drum patterns with high rhythmic fidelity.
This framing thus aligns with the functional and acoustic relationship between drum and VP sounds.
By using drum sounds directly as input, the model can use acoustic features such as stroke intensity and decay characteristics of drum instruments (e.g., snare drum and cymbal).

We formally define the drum-to-VP sound conversion task as follows:
\begin{definition}[Drum-to-VP sound conversion task]
Let $x \in \mathbb{R}^T$ be a drum audio signal of length $T$, and $y \in \mathbb{R}^{T'}$ be a corresponding VP signal of length $T'$.
The task is to learn a function $f:\mathbb{R}^T \rightarrow \mathbb{R}^{T'}$  
that maps $x$ to $y$ preserving perceptual correspondence and rhythmic structure.
\end{definition}

To characterize the objectives of this task, we define three core requirements:
\begin{enumerate}
    \item \textbf{Rhythmic fidelity}: As discussed in Section~\ref{sec:relatedworks}, VP is expected to fulfill a rhythmic function comparable to that of drums in ensemble contexts.  
    Hence, the converted VP sounds should preserve the rhythmic patterns of the input drum sounds.
    \item \textbf{Timbral consistency}: VP sounds often correspond to specific drum instruments and are notated using drum notation or adapted phonetic symbols (e.g., \cite{book}).
    The conversion should thus establish consistent one-to-one mappings from each drum instrument sound to a corresponding VP sounds.

    \item \textbf{Naturalness as VP}: VP is fundamentally produced via human vocal articulation and should include human-like acoustic features, such as aspirated consonants or glottal gestures. The converted output should reflect these characteristics, distinguishing it from purely drum sounds. This requirement corresponds to the preservation of the \textit{VP sound texture}, the perceptual qualities that make a sound recognizable as human-produced VP rather than synthesized drum audio.
\end{enumerate}

\subsection{Design of Subjective Evaluation Criteria}\label{sec:eval}
To evaluate the quality of drum-to-VP conversion, we define subjective evaluation criteria that correspond to the three requirements discussed in \cref{sec:req}. Each criterion is formulated as a binary question to simplify the assessment process in listening tests.
In the following, we summarize the evaluation items and corresponding questions.
For simplicity, we assume that ``Source 1'' and ``Source 2'' are drum input and VP output, respectively.
\begin{enumerate} 
    \item \textbf{Rhythmic fidelity}:
    This criterion evaluates whether the rhythmic structure of the drum input is preserved in the VP output. The corresponding question can be phrased as: \textit{``With regard to rhythmic fidelity, does Source 2 exhibit any unintended change in rhythm compared to Source 1?''}
    \item \textbf{Timbral consistency}: This criterion evaluates whether the each drum instrument is consistently mapped to a corresponding VP sound. The corresponding question can be phrased as \textit{``Regarding timbral consistency, does Source 2 correctly reflect the instrument-to-VP mapping---for example, is the bass drum rendered like a VP bass drum, and the snare drum like a VP snare drum?''}.
    \item \textbf{Naturalness as VP}: This criterion measures whether the converted output is perceptually plausible as a human-produced VP sound. The corresponding question can be phrased as \textit{``Regarding naturalness as VP, does Source 2 sound closer to a drum sound, or to genuine human-produced vocal percussion?''}.
\end{enumerate}
These evaluation criteria guide to design the subjective experiments for drum-to-VP sound conversion, as we will show later in \cref{sec:exp}.


\section{Baseline Methods}
In this section, we first introduce a neural audio synthesis model called RAVE, and then present RAVE-based baselines for drum-to-VP conversion.

\subsection{RAVE \cite{RAVE}}
RAVE, a neural audio synthesis model based on the variational autoencoder (VAE)\cite{Kingma2014ICLR}, was originally developed for musical instrumental sound synthesis and supports real-time audio generation.
It reconstructs the relationship between a $D$-dimensional input signal $\bm{x}\in\R^{D}$ and a $H$-dimensional latent variable $\bm{z}\in\R^{H}$ via neural networks.
The decoder maps $\bm{z}$ back to $\bm{x}$. 
To preserve rhythmic characteristics, it is designed to output an amplitude envelope separately, which is applied via element-wise multiplication to shape the resultant waveform.

The training objective maximizes the variational lower bound on the log-likelihood of the data. This is achieved by jointly optimizing the reconstruction loss between $\bm{x}$ and its estimate from $\bm{z}$, and the Kullback--Leibler (KL) divergence between the approximate posterior $q(\bm{z} \mid \bm{x})$ and the prior $p(\bm{z})$, typically assumed to be standard normal distribution.

One variant of RAVE employs vector quantized VAE (VQ-VAE) \cite{vandenOord2017NIPS}, a method that discretizes the latent space using vector quantization.
This version is partially incorporated into the official RAVE implementation\footnote{\url{https://github.com/acids-ircam/RAVE}}, although it is not mentioned in the original RAVE paper \cite{RAVE}.

The learning procedure of RAVE consists of two stages. In the first stage, the encoder and decoder are trained jointly using the VAE objective. The reconstruction loss is computed in the magnitude spectrogram domain using a short-time Fourier transforms (STFT) with multiple time–frequency resolutions \cite{Engel}, which captures audio characteristics across multiple time-frequency resolutions.
In the second stage, RAVE uses adversarial training to further refine the decoder. The encoder is fixed, and the decoder is further optimized using a generative adversarial network (GAN) framework \cite{Goodfellow2014NIPS}. Here, the decoder acts as the generator, and the loss function is a weighted sum of an adversarial loss, a multi-scale spectral loss~\cite{Yamamoto2020ICASSP}, and a feature-matching loss \cite{Kumar}. The official RAVE implementation utilizes multi-period~\cite{Kong2020NEURIPS} and multi-scale discriminators~\cite{Wang2018CVPR}, which operate at various temporal resolutions to improve audio quality, in contrast to the single-discriminator approach based on hinge loss described in the original literature~\cite{RAVE}.


\subsection{Baseline Methods}
\begin{figure}[t]
    \centering
    \begin{subfigure}[b]{\hsize}
        \centering
        \includegraphics[width=0.9\linewidth]{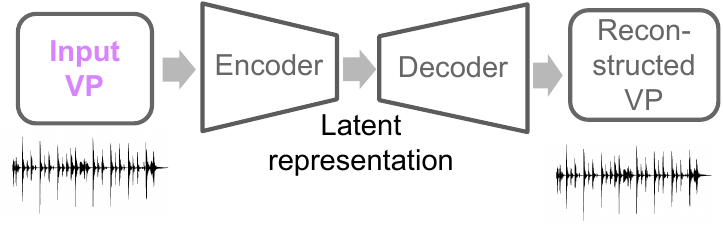}
        \vspace{-4mm}
        \caption{Training stage}
        \vspace{2mm}
    \end{subfigure}
    \begin{subfigure}[b]{\hsize}
    \centering
        \includegraphics[width=0.9\linewidth]{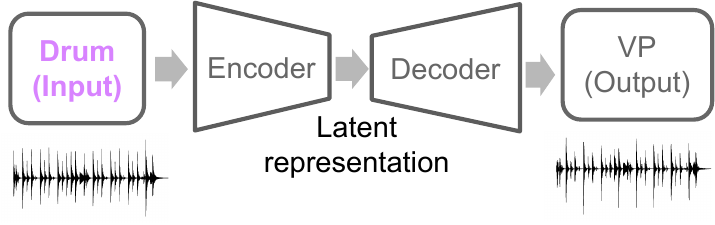}
        \vspace{-4mm}
        \caption{Conversion stage}
        \vspace{2mm}
    \end{subfigure}
    \caption{Workflow of the RAVE-based baselines.
    (a) During training, the model is trained on a VP dataset. (b) At the inference time, the trained model is applied to drum sounds to generate the corresponding VP sounds.}
    \label{fig:proposed}
\end{figure}
We construct drum-to-VP sound conversion baselines using RAVE.
RAVE also supports timbre transfer: when trained on audio of a particular instrument, it can generate output that reflects the timbral characteristics of the training data while preserving the input rhythm.
We leverage this capability by training RAVE on a VP dataset and applying it to drum sounds (see \cref{fig:proposed}).
We implement two baselines: a VAE-based model (RAVE) and a VQ-VAE-based variant (VQ-RAVE).


RAVE offers an additional advantage in terms of data preparation.
It does not rely on explicitly aligned parallel data between drum and VP sounds, yet its timbral consistency has been validated experimentally.  
This capability makes RAVE suited for settings where aligned drum--VP datasets are unavailable.
For example, the jaCappella corpus~\cite{nakamura2023jacappella} contains only VP recordings and lacks paired drum audio or symbolic alignment.
Our approach circumvents this limitation, making it broadly applicable.

\section{Subjective Evaluation Experiment}\label{sec:exp}
\subsection{Training of Baseline Models}
A subjective evaluation experiment was conducted to assess the performance of the baseline methods.

\noindent \textbf{Dataset}: We used the jaCappella corpus \cite{nakamura2023jacappella}, a Japanese contemporary a cappella dataset. The corpus consists of 10 subsets, each containing five songs arranged for six distinct vocal parts, including VP. Each voice part is available as a separate monaural audio track.
Following the official training/test data split, one song was selected from each subset for use as the validation data. The remaining 40 songs were used as training data. All audio signals were resampled to \qty{44.1}{kHz} and converted to monaural format.

\noindent \textbf{Preprocessing}: Preprocessing follows the official implementation of RAVE.
It involved silence-based segmentation using the \texttt{pydub} library, with silence defined as intervals below \qty{-60}{dBFS} for more than one second. Data augmentation techniques included random gain, random muting of segments, and dynamic range compression using the \texttt{sox} audio processor.

\noindent \textbf{Model configuration}: Two models were trained: RAVE and VQ-RAVE. 
Both employed causal convolution filters for real-time compatibility. 
The implementations used the official RAVE repository,
and hyperparameters were set according to its default settings, as specified in the configuration files\footnote{\url{https://github.com/nbrnk/RAVE}}.
Models were trained for $300{,}000$ epochs using the Adam optimizer with a learning rate of \num{1.0e-3} and momentum parameters of $0.5$ and $0.9$. Discriminators used in adversarial training were also optimized with Adam with a learning rate of \num{1.0e-4}.

\subsection{Experimental Setup for Subjective Evaluation}\label{sec:test_data}
\begin{figure}[t]
    \begin{subfigure}[b]{\hsize}
        \includegraphics[width=\linewidth]{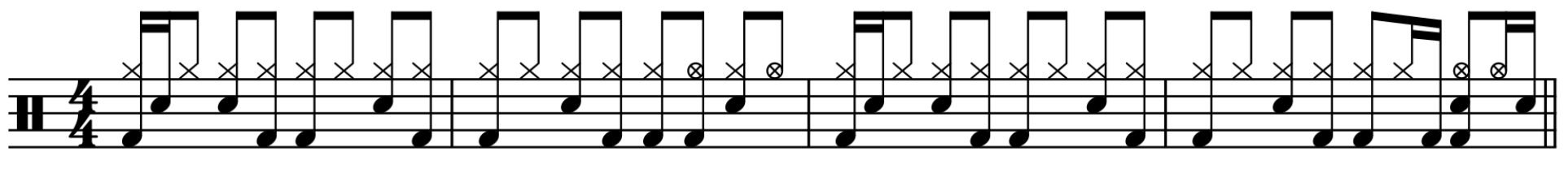}
        \vspace{-6mm}
        \caption{First drum pattern}
        \vspace{2mm}
    \end{subfigure}
    \begin{subfigure}[b]{\hsize}
        \includegraphics[width=\linewidth]{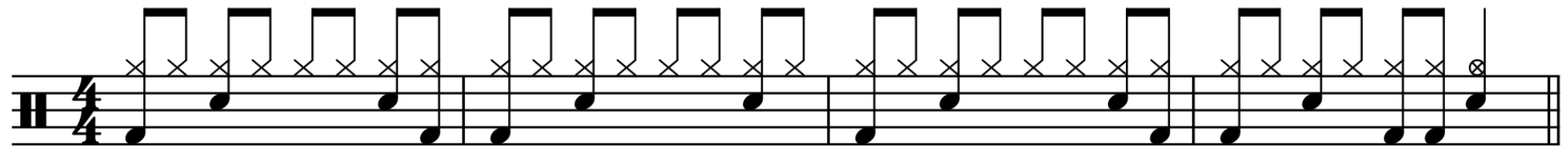}
        \vspace{-6mm}
        \caption{Second drum pattern}
        \vspace{2mm}
    \end{subfigure}
    \begin{subfigure}[b]{\hsize}
        \includegraphics[width=\linewidth]{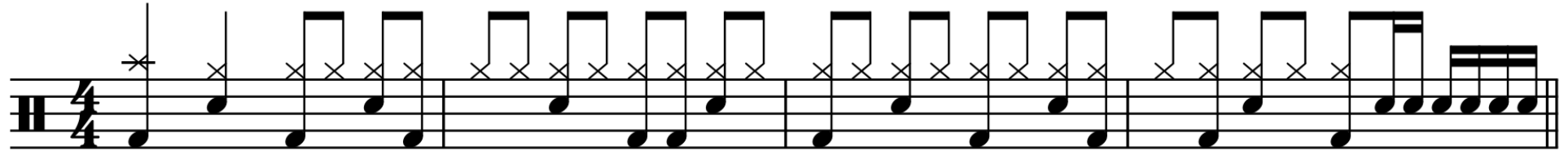}
        \vspace{-5mm}
        \caption{Third drum pattern}
        \vspace{2mm}
    \end{subfigure}
    \caption{Standard drum notation of 3 drum patterns for subjective evaluation.}
    \label{fig:drums}
\end{figure}

The test drum audio was synthesized using a commercial virtual drum instrument, Ezdrummer~3\footnote{\url{https://www.toontrack.com/product/ezdrummer-3/}}. 
We used the default acoustic drum kit.
\Cref{fig:drums} shows the drum notation used.
Three different patterns were selected, each at three tempos (80, 120, 160 beats per minute (BPM)), yielding 9 test cases.
To ensure sufficient temporal length for the development of drum patterns, the test data were set to a duration of 4 measures in 4/4 time, corresponding to approximately 6–12 seconds.
The patterns were selected to avoid simultaneous triggering of multiple drum instruments as far as possible, aligning with the monophonic nature of VP performance.



\begin{figure}[t]
\centering
\includegraphics[width=0.9\columnwidth]{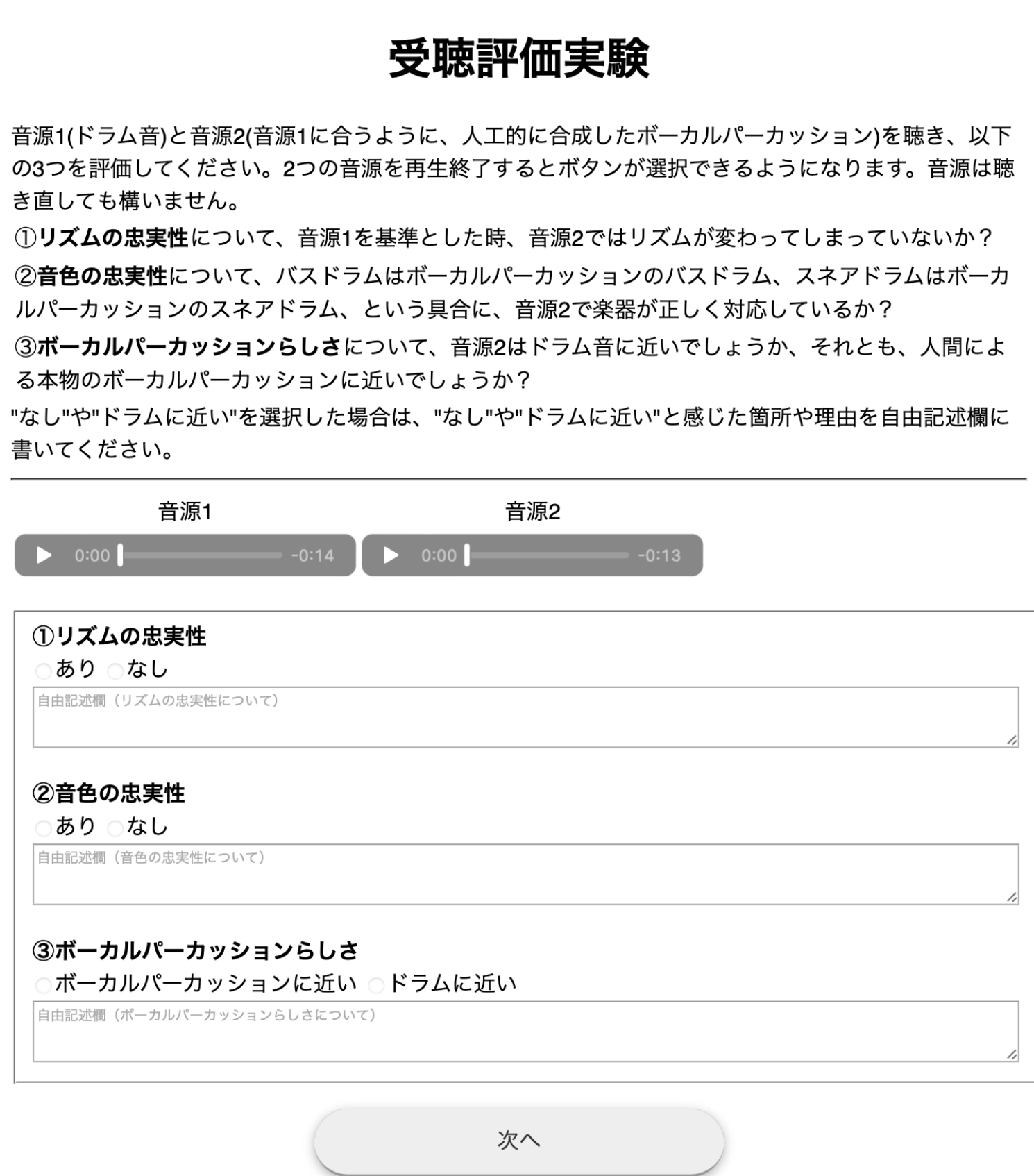}
\caption{
    Web interface used for the subjective evaluation experiment.
    The instructions were written in Japanese.
    The three evaluation questions were exactly those defined in \cref{sec:eval}.
    An English translation of the remaining instructions is as follows: \textit{``Please listen to Source 1 (drum audio) and Source 2 (artificially synthesized vocal percussion designed to match Source 1), and evaluate them based on the following three criteria. The answer buttons will become available after both sources have finished playing. You may replay the audio if needed. If you select ``None'' or ``Closer to drum'' please describe the part or reason that led you to feel that way in the free-comment box.''}
}
\label{fig:web_interface}
\end{figure}

We recruited 6 Japanese participants (male and female, aged in their 20s to 40s) via the crowdsourcing platform Lancers\footnote{\url{https://www.lancers.jp}}, restricting participation to individuals with prior VP experience to ensure evaluative reliability.
Each participant evaluated 18 audio pairs: one set using input drum sounds and their corresponding RAVE outputs, and the other using input drum sounds and their corresponding VQ-RAVE outputs.
To conduct the evaluation, we implemented a web-based interface that allowed participants to freely replay both the drum and VP sounds as many times as needed.
The interface was designed in accordance with the criteria and question formulations described in \cref{sec:eval}.
Each audio pair had to be played at least once before proceeding, and participants answered the three questions for rhythmic fidelity, timbral consistency, and naturalness as VP.
\Cref{fig:web_interface} shows a screenshot of the evaluation interface.

\subsection{Results}
\begin{figure}[t]
    \begin{minipage}[t]{\hsize}
    \centering
    \includegraphics[width=.75\linewidth]{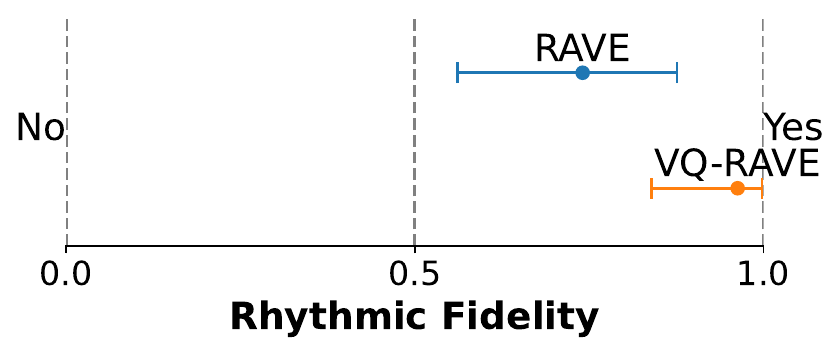}
    \end{minipage}
    \begin{minipage}[t]{\hsize}
    \centering
    \includegraphics[width=.75\linewidth]{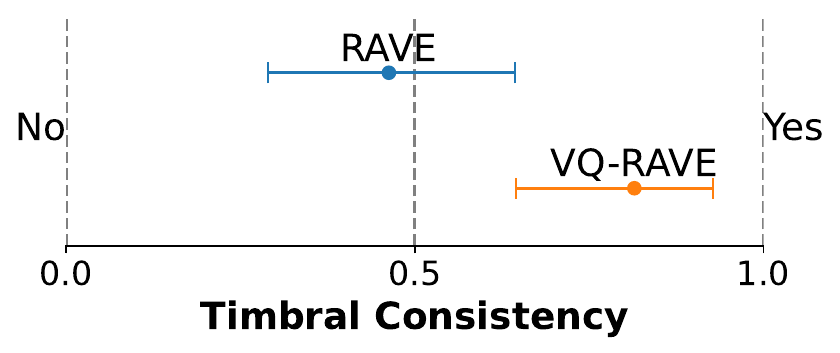}
    \end{minipage}  
    \begin{minipage}[t]{\hsize}
    \includegraphics[width=\linewidth]{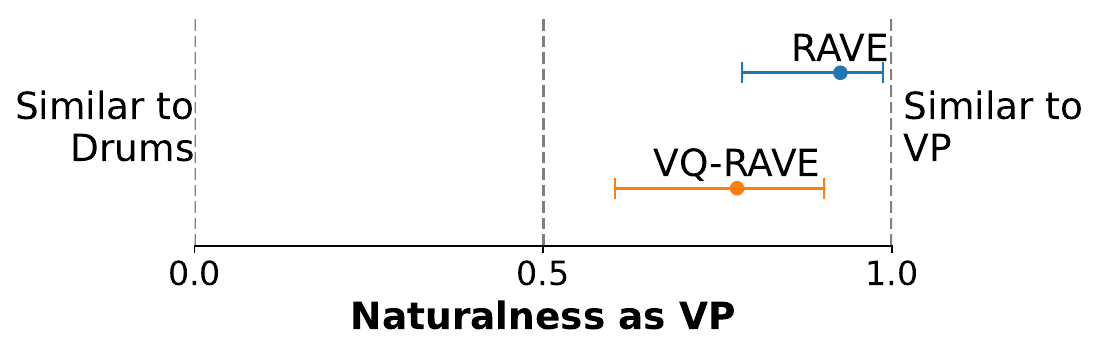}
    \end{minipage}
    \caption{Average subjective evaluation scores with 99\% confidence intervals for all criteria.}
    \label{fig:result}
\end{figure}

\Cref{fig:result} shows the evaluation results obtained from the listening tests. Binary scores were assigned based on participant responses: a score of 1 was given when participants judged that the rhythm was maintained, the mapping between instruments and VP sounds was consistent, and the output sounded similar to human-produced vocal percussion, corresponding to rhythmic fidelity, timbral consistency, and naturalness as VP, respectively. All other responses were assigned a score of 0.
The 99\% confidence intervals were obtained using the Clopper--Pearson method~\cite{clopper1934use}, which accounts for the binary nature of the responses and ensures bounds within [0, 1].

RAVE achieved scores that were statistically significantly higher than the chance level (0.5) in rhythmic fidelity and naturalness, but not in timbral consistency.
VQ-RAVE yielded significant results across all three criteria, suggesting more consistent and structured mappings.
These results demonstrate that both methods can produce intelligible and rhythmically aligned VP sounds, validating their suitability as baselines for the drum-to-VP sound conversion task.

The gap between RAVE and VQ-RAVE appears to be linked to the latent representation: RAVE uses continuous variables, whereas VQ-RAVE employs discrete codes. As discussed in \cref{sec:relatedworks}, VP sounds may benefit from symbolic or categorical modeling (e.g., IPA), potentially explaining the advantage of VQ-RAVE in timbral control.
However, the higher rating of RAVE in terms of naturalness may indicate that its outputs, while less structured, sounded more plausibly human-produced. This qualitative observation is further explored in the next section.


\subsection{Analysis Based on Free-Text Comments}
We analyzed the participant comments collected through the free-text forms. Since all participants were Japanese, the comments shown below were translated into English by the first author. For RAVE, most comments focused on the resonance of hi-hats and cymbals, as well as the breathy quality inherent in VP. Representative comments included:
\begin{itemize}
    \item ``The metallic resonance of hi-hats and cymbals is well captured.''
    \item ``Too much air noise, so it sounds like VP.''
    \item ``Lacks bass drum weight, sounds mouth-produced.''
    \item ``Poor timbral consistency; snare drum and cymbals are especially weak.''
\end{itemize}
For VQ-RAVE, most comments highlighted its balance between instrument-like clarity and vocal percussion expressiveness, especially for snare drums. However, cymbals were often described as overly metallic and drum-like. Representative comments included:
\begin{itemize}
    \item ``Snare drum is well reproduced.''
    \item ``Distinctions between musical instruments are clearer; still sounds like VP.''
    \item ``Tom-like snare drums differ from real drums but are expressive as VP.''
    \item ``Snare drum sounds inward and VP-like.''
    \item ``Cymbals are too metallic, like real drums.''
\end{itemize}

Notably, reactions to the naturalness item varied: some saw the deviation from drum realism as a virtue of VP synthesis, while others found it too artificial. This ambiguity suggests that ``naturalness'' in VP synthesis spans a spectrum from realism to stylization, and future questionnaires should better capture this range.

\section{Conclusion}
We presented a drum-to-VP sound conversion task, conceptualized as a timbre transfer problem rather than a traditional speech synthesis task.  
This formulation aligns with the nature of VP, which imitates percussive instruments through vocal articulation rather than conveying linguistic content. To properly evaluate synthesized VP sounds, we defined three core requirements for successful conversion: rhythmic fidelity, timbral consistency, and naturalness as VP. These criteria served as guiding principles for both model development and evaluation. To establish baselines for this task, we implemented two RAVE-based systems: one with continuous latent variables and the other with vector quantization. Subjective evaluations showed that both models successfully performed drum-to-VP conversion, with the VQ-based system yielding more consistent mappings and higher perceptual scores. This work lays a foundation for future research on VP synthesis by establishing a clear task definition, baseline systems, and evaluation protocol.

\section*{Acknowledgment}
This work was supported by JSPS Grants-in-Aid for Scientific Research JP23K18474, JP21H04900, JP23K28108, JST Fusion Oriented REsearch for disruptive Science and Technology (FOREST) JPMJFR226V and Tateisi Science and Technology Foundation.

\printbibliography

\end{document}